\documentclass[prb,num, preprint, showpacs, 10pt, twocolumn,superscriptaddress]{revtex4-2}
\usepackage{graphicx}
\usepackage[utf8]{inputenc}
\usepackage[colorlinks=true, pdfstartview=FitV, linkcolor=blue, citecolor=blue]{hyperref}
\usepackage{color}
\usepackage{colortbl}
\usepackage{epstopdf}
\usepackage{graphicx}
\usepackage{amsmath}
\usepackage{array,multirow}
\usepackage[table]{xcolor}% http://ctan.org/pkg/xcolor
\usepackage{tabularx}
\usepackage{float}
\usepackage{ctable}
\makeatletter
\def\hlinewd#1{%
\noalign{\ifnum0=`}\fi\hrule \@height #1 %
\futurelet\reserved@a\@xhline}

\begin{document}

% Use the \preprint command to place your local institutional report number 
% on the title page in preprint mode.
% Multiple \preprint commands are allowed.

\preprint{}

\title{Strain-Induced Curvature in Monolayer Graphene: Effects on Electronic Structure, Phonon Dynamics, and Lattice Thermal Conductivity}

\author{M. C. Santos}
\affiliation{Future Technology School, Shenzhen Technology University, Shenzhen 518118, P.R. China}

\author{E. Lora da Silva}
\email{estelina.silva@ua.pt}
\affiliation{IFIMUP, Institute of Physics for Advanced Materials, Nanotechnology and Photonics, Departamento de F\'{i}sica e Astronomia da Faculdade de Ci\^{e}ncias da Universidade do Porto, Rua do Campo Alegre s/n, 4169-007 Porto, Portugal}
\affiliation{High Performance Computing Chair, University of Évora, Rua Romão Ramalho 59, 7000-671 Évora, Portugal}
\affiliation{Future Technology School, Shenzhen Technology University, Shenzhen 518118, P.R. China}

\author{D. S. Baptista}
\affiliation{IFIMUP, Institute of Physics for Advanced Materials, Nanotechnology and Photonics, Departamento de F\'{i}´sica e Astronomia da Faculdade de Ci\^{e}ncias da Universidade do Porto, Rua do Campo Alegre s/n, 4169-007 Porto, Portugal}

\author{T. Santos}
\affiliation{University of Aveiro, TEMA – Centre for Mechanical Technology and Automation, Campus Universitário de Santiago, 3810-193 Aveiro, Portugal}

\author{M. Molinari}
\affiliation{Department of Physical and Life Sciences, University of Huddersfield, Huddersfield HD1 3DH, United Kingdom}

\author{F. J. Manj\'{o}n}
\affiliation{Instituto de Dise\~{n}o para la Fabricaci\'{o}n y Producci\'{o}n Automatizada, MALTA Consolider Team, Universitat Polit\`{e}cnica de Val\`{e}ncia, Val\`{e}ncia, Spain}

\author{Yin Cui}
\affiliation{Future Technology School, Shenzhen Technology University, Shenzhen 518118, P.R. China}

\author{Xidong Lin}
\affiliation{Future Technology School, Shenzhen Technology University, Shenzhen 518118, P.R. China}

\author{Tao Yang}
\email{yangtao@sztu.edu.cn}
\affiliation{Future Technology School, Shenzhen Technology University, Shenzhen 518118, P.R. China}

\begin{abstract}
  We present a comprehensive set of calculations to investigate the effect of strain-induced x-y topological perturbation in the monolayer graphene sheet. We show that the induced curvature with the defined strain constraint, energetically stabilizes the systems. The electronic properties are modified when the amplitude of the curvature of the sheet increases, which induces Van Hove singularities of the electronic Density of States to approach the Fermi energy. The highly curved system exhibits coexisting flat and linear dispersions close to the Fermi level, which is a promising feature for thermoelectric applications. We also demonstrate, through the phonon dispersion curves, that respective systems are dynamically stable within the studied range of strains/curvatures. Moreover, the flexural acoustic mode transitions from quadratic to linear dispersion under strain, mimicking the 3D behavior and enhancing phonon scattering. The increase of phonon scattering will therefore decrease the value of the lattice thermal conductivity, $\kappa_L$.  Such results allows us to conclude that it is possible to tune $\kappa_L$ by applying x-y strain to the monolayer sheet, and inducing different topological curvatures. 
\end{abstract}

\maketitle %\maketitle must follow title, authors, abstract and \pacs

\date{\today}

%%%%%%%%%%%%%%%%%%%%%%%%%%%%%%%%%%%%%%%%%%%%%%%%%%%%%%%%%%%%%%%%%%%%%
%% Start the main part of the manuscript here.
%%%%%%%%%%%%%%%%%%%%%%%%%%%%%%%%%%%%%%%%%%%%%%%%%%%%%%%%%%%%%%%%%%%%%
\section{Introduction}

Graphene is a two dimensional (2D) material embedded in three dimensions (3D) space, since one should expect deviations from a true flat surface due to thermal fluctuations that perturb the system. Structural corrugations and ripples, have been observed in suspended graphene, and furthermore, atomistic simulations have also shown that ripples appear spontaneously due to thermal fluctuations \cite{JPhysCondMatt.26.18.2014, RevModPhys.81.109.2009}.

Geometric curvature and strain-induced deformations in graphene can give rise to pseudo-magnetic fields, which can lead to observable phenomena, such as Landau levels at the density of states \cite{Science.329.544.2010}. Moreover, such deformations impact the energy spectra (i.e. tuning of the electronic band gap), particularly under magnetic and electrical fields, and can enable strain-engineered quantum transport \cite{10.1039/9781788015882-00287,SILVA2020373}.

Graphene-type materials are inevitably subject to out-of-plane deformations, as described by the physics of the Dirac oscillator equation \cite{SANTOS2021168429, JPhysCondMatt.26.18.2014}. As a consequence of the Dirac harmonic oscillator equation, it is possible to address the physics of the Landau levels and also consider the theoretical framework of Jaynes-Cummings, and the relevant optical properties of curved graphene \cite{SANTOS2021168429, JPhysCondMatt.26.18.2014}.

%### Core Connection: Why Flat Bands for Thermoelectrics?

Thermoelectric materials convert heat (temperature gradients) directly into electricity and vice versa. Respective efficiency is defined by the dimensionless figure of merit, the Seebeck coefficient ($S$), where the challenge to obtain a good thermoelectric material is to try and optimize the interdependent parameters that make up $S$. One of these parameters is the engineering of the room temperature lattice thermal conductivity, $\kappa_L$ \cite{PhysRevB.104.024103,HUSSAIN2025109672, ma15238672,WOOD2023250, doi:10.1021/acs.chemmater.7b02253}, which can be modified by applying external perturbation fields, such as strain, or through defect engineering.

Van Hove Singularities (VHS) are sharp peaks or divergences occurring in the electronic density of states (eDoS), or occur at points where the dispersion relation (energy as a function of momentum) becomes flat. These are observed mainly at the low energy regime. It is known that VHS, mainly occur in twisted bilayer graphene (tBLG), and also for 2D transition metal dichalcogenides. For tBLG, and by tuning the twist angles, the VHS can be positioned closer to the Fermi energy (E$_F$), therefore inducing strong electron correlations, superconductivity, magnetism and tuneable electronic phases crucial for next-generation of thermoelectric and quantum devices \cite{Nature_556_43, 1t3m-2sjn}.

Flat electronic dispersion bands correspond to heavy charge carriers (large effective masses, m$^*$), and are a playground for exploring correlated electron physics, which not only include superconductivity, but also efficient thermoelectricity \cite{1t3m-2sjn}. For the latter, these flat bands (or a high eDoS, close to the E$_F$), tends to enhance $S$, since it alters $\kappa_L$ by affecting the phonon scattering. These are often associated with localized states and low mobility; however under certain conditions of symmetry, geometry, or interaction it is possible to observe simultaneously both flat and dispersive bands (light charge carriers), close to the Fermi energy, along different high-symmetry directions of reciprocal space \cite{PhysRevMaterials.9.L032401,doi:10.1021/acs.jpcc.4c08588,MADSEN2018140}. 

%In spite this, many materials hosting flat bands (e.g., complex crystals, layered materials, "cage-like" structures) inherently have strong anharmonicity or point defects, and thus can also suppress $\kappa_L$ due to scattering effects, and thus may evidence short lifetimes of the charge carriers \cite{PhysRevMaterials.9.L032401,doi:10.1021/acs.jpcc.4c08588,MADSEN2018140}.

Ideally, and to ensure an efficient $\kappa_L$, one should expect a flat-and-dispersive electronic band structure \cite{PhysRevMaterials.9.L032401,doi:10.1021/acs.jpcc.4c08588,MADSEN2018140}, where the band is flat along the direction closer to the Fermi energy, and dispersive along other directions to maintain conductivity. 

Gunst \textit{et al.}~\cite{1t3m-2sjn} conducted a theoretical study to analyze the thermoelectric properties of tBLG. They observed that through respective thermoelectric properties, the flat bands and their tunability lead to giant, gate-tunable $\kappa_L$ \cite{1t3m-2sjn}.

Moreover, from the work of Cheng \textit{et al.} \cite{CHENG2023101093} the authors observed a local minimum in thermal conductivity in tBLG at the 'magic' angle of 1.08°. Within the supercell of a Moiré lattice, different stacking modes generate phonon scattering sites which reduce the thermal conductivity. The thermal magic angle would rise from the competition between the delocalization of atomic vibrational amplitudes and stresses, which  weakens the scattering strength of a single scatterer. On the other hand, the increased AA stacking density also increases the density of scatterers. The combination of the two effects eventually leads to irregularity in heat conduction, where the manifestation of a magic angle, can evidence novel low dimensional thermal mechanisms.

In this work we study, through \textit{ab-initio} methods, the monolayer graphene sheet where strain is induced, and thus deformation is tuned on the monolayer surface. We show that the induced curvature with the defined strain constraint, energetically stabilizes the systems, being more stable than flat graphene when the same strain magnitude is applied. We also show by calculating the phonon dispersion, that the studied systems are dynamically stable within the range of analyzed strains/curvatures. Since the electronic properties modify, as the deformation increases, we observe that it is possible to tune $\kappa_L$ with different curvature topology, since the VHS can be adjusted closer to the E$_F$ when deformation increases. Moreover, from the phonon band dispersion, we observe that the acoustic flexural mode changes proportionally as strain and deformation increases, modifying from a quadratic-type feature towards a linear dispersion character, and with similar characteristics to a 3D material.

\section{Theoretical Methodology}

Structural and electronic properties were calculated within the framework of density-functional theory (DFT)\cite{hohenberg-pr-136-1964}. The Vienna \textit{Ab-initio} Simulation Package (VASP - v.6.4.3) code \cite{kresse-cms-6-1996} was employed to perform simulations with the projector augmented-wave (PAW) scheme, which included C[2s$^2$2p$^2$] as valence electrons. Convergence of the total energy was achieved with a plane-wave kinetic-energy cut-off of 600 eV. The generalized-gradient approximation (GGA) functional with the Perdew-Burke-Ernzerhof parameterization revised for solids (PBEsol) \cite{perdew-prl-100-2008,perdew-prl-102-2009}, was used for all the calculations. 

The Brillouin-zone (BZ) was sampled with $\Gamma$-centered Monkhorst-Pack \cite{monkhorst-prb-13-1976} grids employing an adequate density mesh of 11$\times$11$\times$1, for the  5$\times$5$\times$1 system. For the electronic Density of states (eDoS), and to refine the BZ for precision, we increased these points up to 41$\times$41$\times$1.

The vacuum region was defined at 15 \AA~for all the studied systems. Atomic positions were fully optimized to obtain relaxed structures at a range of fixed surface-areas (by reducing the scaling parameter one obtains a higher amplitude of curvature and strain), in order to conserve the intended curvature amplitude.  

The phonon properties were computed by using the supercell finite-displacement method implemented in the Phonopy package \cite{togo-prb-78-2008}, with VASP used as the force calculator. 'Supercells' were kept with its original size, as a 5$\times$5$\times$1 cell, due to the large lattice parameters of the system. This procedure allowed for the exact calculation of frequencies at the zone center ($\Gamma$) and inequivalent zone-boundary wavevectors, which were then interpolated to obtain the phonon dispersion curves. 

Lattice thermal conductivity ($\kappa_\textit{L}$) was computed by resorting to the Phono3py software \cite{chaput-prb-84-2001, PhysRevB.91.094306.2015}; and VASP, once again, was used as the calculator to obtain the third-order interatomic force constants. The original supercell cell size was considered with a \textbf{q} mesh of 6$\times$6$\times$1 and resorting to the tetrahedron method to perform the integration for the phonon lifetime calculation. The phonon lifetimes were computed with the single-mode relaxation-time approximation, to solve the Boltzmann transport equations \cite{chaput-prb-84-2001, PhysRevB.91.094306.2015}.

\section{Results and Discussion}

\subsection{Structural and Energetic Stability}
\label{energy_stability}

\begin{figure*}
\begin{center}
\includegraphics[width=14cm]{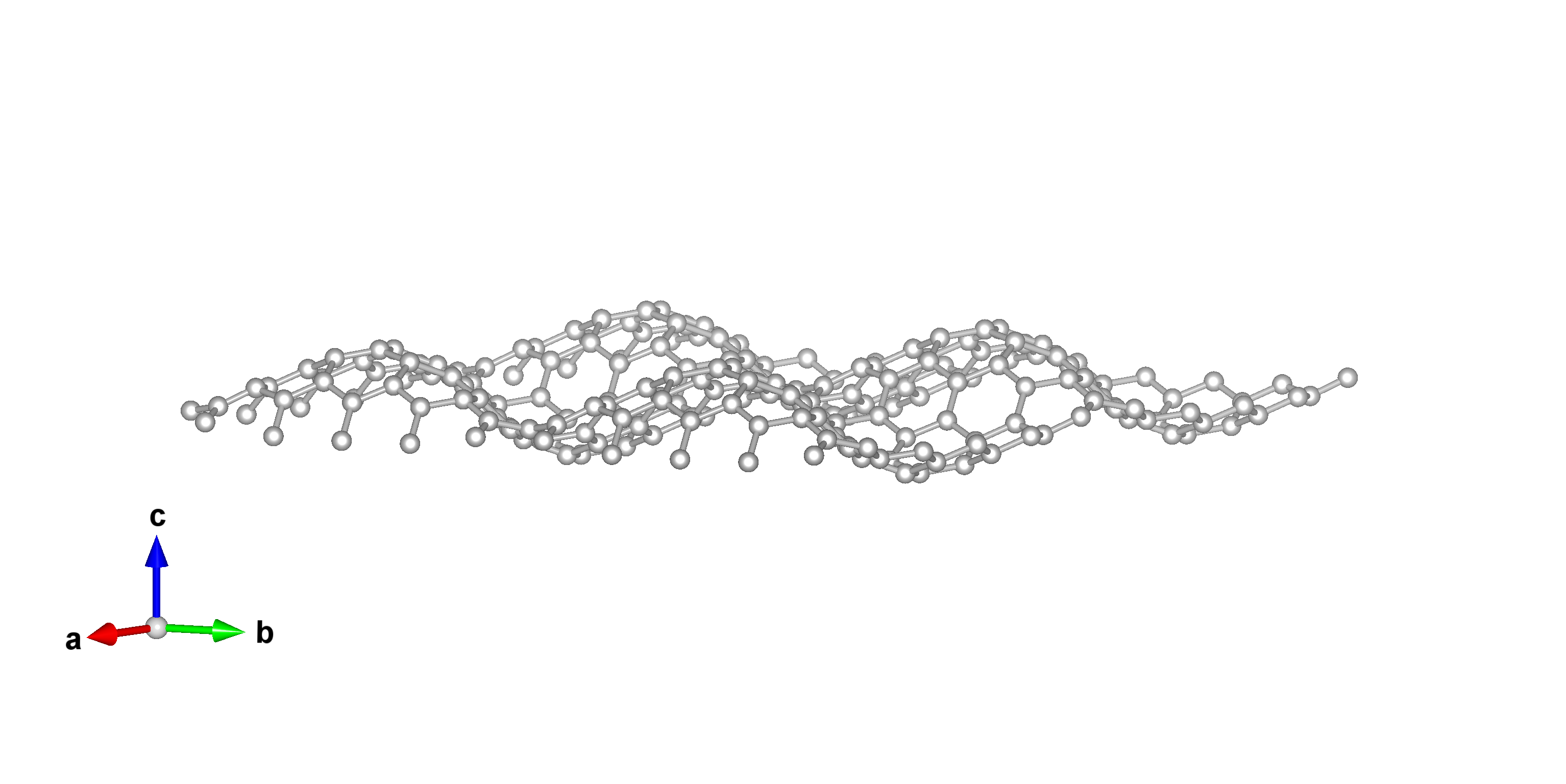}
\caption{\label{fig:structure}
The 5$\times$5$\times$1 cell with highest amplitude deformation and x-y induced strain (S5), and yielding the \textit{P3m1} (S.G.156) space group symmetry.} 
\end{center}
\end{figure*}

We have performed calculations on a series of 5$\times$5$\times$1 cells (Figure \ref{fig:structure}), with different induced strain along the x-y directions (Table \ref{table:deform}), to obtain topological deformations with minimum of applied strain. This procedure allowed us to obtain different deformation amplitudes of the lattices, and therefore different electronic and phonon properties, which will be detailed in the following subsections.

Such calculations were carried out by constraining respective lattice parameters and slowly distancing two specific C atoms [located at C$_1$=(1/3, 2/3, 0) and C$_2$=(2/3, 1/3, 0) of the 5$\times$5$\times$1 cell] away from the z-direction (inverse directions), thus leading to the required deformation after the atomic relaxations were performed (Table \ref{table:deform}). The deformation parameter, D$_\textrm{ef}$, is obtained by multiplying the amplitude of the curvature with the frequency of the unit-cell of the strained lattice.

\begin{table}[h]
\small
  \caption{Structural parameters of the different graphene 5$\times$5$\times$1 supercells (\textit{P3m1}, S.G. 156), with compressed lattice parameters ($\Delta$ a$_0$) and different deformation values, D$_\textrm{ef}$ (from S1 to S5). $\Delta$ a$_0$ is the lattice paramter relative to the planar monolayer supercell (M) where a$_0$=12.23~\AA.}
  \label{table:deform}
  \begin{tabular}{ccc}
    \hline
 ID &  $\Delta$ a$_0$ [\%] &	D$_\textrm{ef}$   \\
    \hline
   M &   0.0  & 	---    \\ \hline
 S1  & -0.0116 & 0.328249 \\ \hline   %3.41 GPa 551_5
 S2 & -0.0318 & 1.288536    \\ \hline %.5.56 GPa 551_6
  S3 & -0.0419 & 1.611308  \\ \hline   %6.94 GPa  551_4
S4& -0.0923 & 2.551689	  \\ \hline   %14.33 GPa  551_2
S5 & -0.1175 & 2.813703    \\ \hline  % 17.63 GPa   551_3
    \hline
  \end{tabular}
\end{table}

\begin{figure}[!]
\begin{center}
\includegraphics[width=8cm]{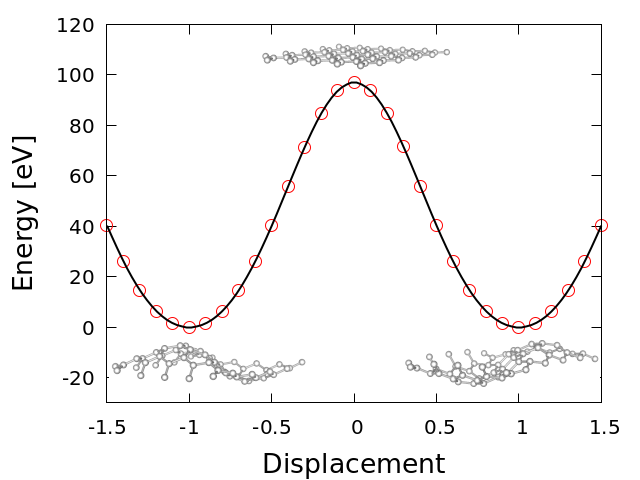}
\caption{\label{fig:well_551}
Potential-energy surface of the S5 graphene system (where a$_0$-0.1175 \%), for which the curves have been normalized with respect to the highest distortion amplitude (energetic stable system).} 
\end{center}
\end{figure}

We find that as we increase the x-y strain, the sheet stabilizes with higher deformation amplitude. To demonstrate that the deformations, with induced strain, are energetically more stable when compared to the well known flat monolayer system, we have analyzed respective stability by calculating the potential-energy surfaces (Figure \ref{fig:well_551}) for the system with largest deformation (S5). 

By interpolating the curved surface onto a planar geometry, and subsequently following the transformation path toward the system with inverse curvature topology, we observe that the potential well of the S5 system is considerably deep. The energy difference between the deformed configuration and the flat surface is approximately 100 eV, indicating that a strained, deformed surface is energetically preferred. With the exception of the system having the lowest amplitude (S1), the potential wells exhibit relatively high energy levels (Figure \ref{fig:potential_well_4}). Consequently, we conclude that curvature represents a robust and stable topology, characterized by a very high potential energy relative to that of flat monolayer graphene.

\subsection{Electronic Properties}

\begin{figure*}[!]
\begin{center}
\includegraphics[width=16cm]{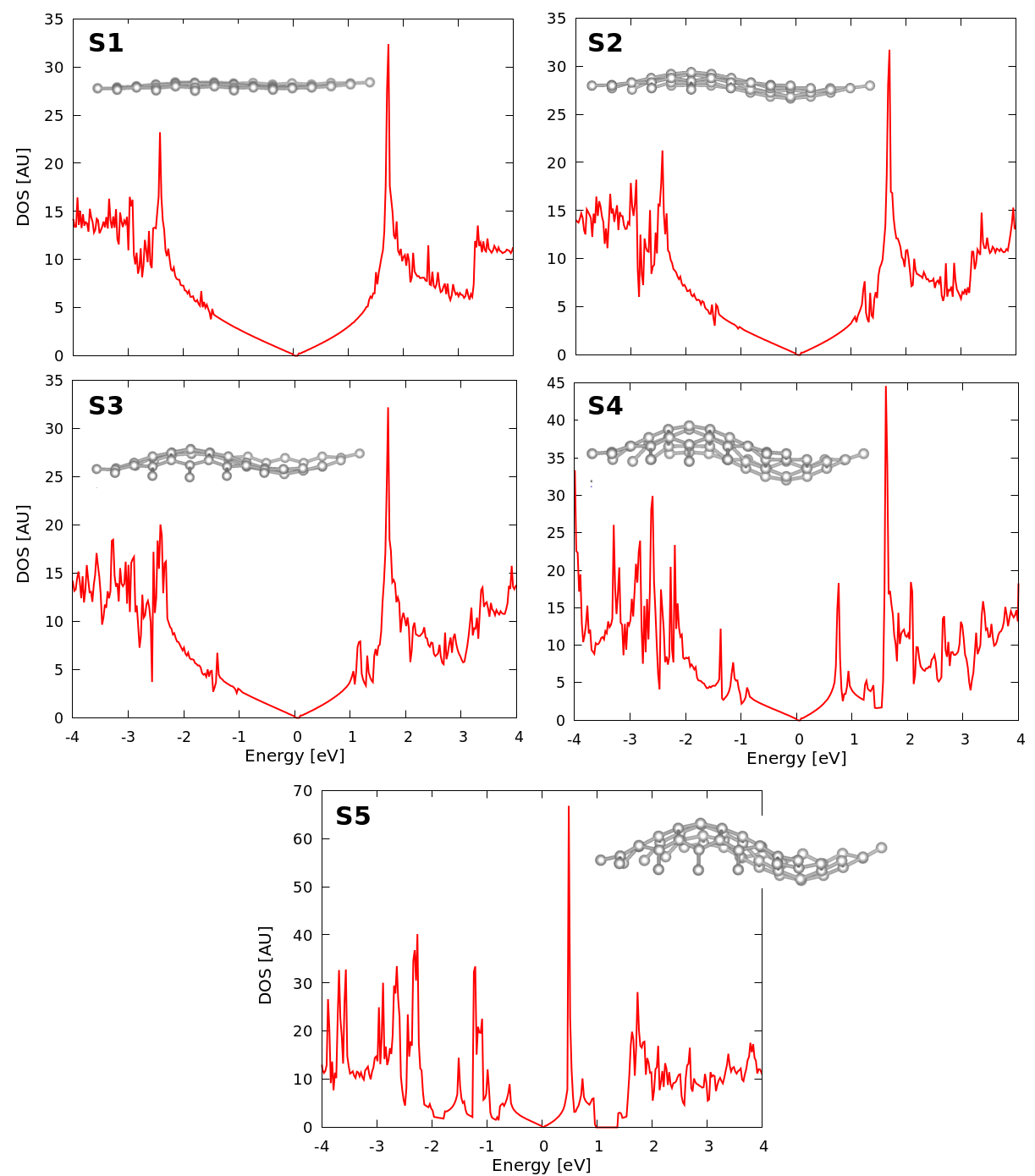}
\caption{\label{fig:elec_dos_551}
Density of States of the 5$\times$5$\times$1 cells with different curvature deformations, from S1 until S5.} 
\end{center}
\end{figure*}

The eDoS have been computed for all the five studied graphene systems describe in Table \ref{table:deform}. The S1 system, showing very similar features as to the flat graphene monolayer, evidences two main VHS \cite{Dav23} (Figure \ref{fig:elec_dos_551}).

Here we observe that one of the VHS is positioned at the valence band, around -2.5 eV, and the second singularity (more resonant) at the conduction band, is very close to 2 eV. As the strain of the lattice increases, hence the amplitudes, we observe significant changes to the eDoS, mainly to the singularities at the conduction band, which increase in magnitude, and approach the E$_F$. For the S5 system, the VHS at the conduction band is very close to 0.5 eV, and is very pronounced. The VHS at the valence bands also change dramatically, when increasing the deformation from the S1 to the S5 system, becoming significantly attenuated for S5.

The sharp eDoS peaks associated with the VHS emerge when the system exhibits strong responses to electron–electron interactions, potentially giving rise to complex and intriguing phenomena. We demonstrate that the curvature topology modifies the VHS, enabling these to approach the Fermi energy as the curvature amplitude increases. Such feature opens intriguing prospects for VHS engineering of the electronic phases of deformed graphene sheets.

\begin{figure*}[!]
\begin{center}
\includegraphics[width=10cm]{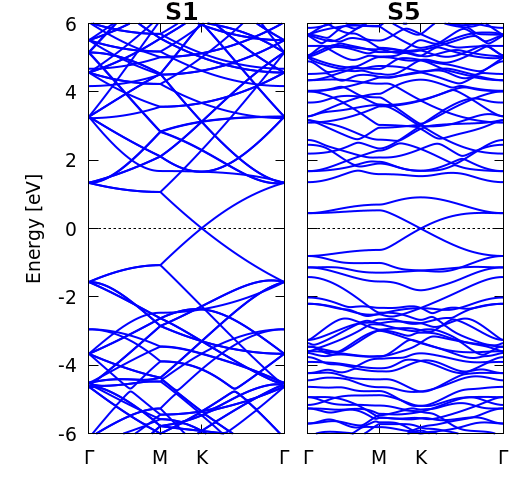}
\caption{\label{fig:elec_band_551}
Electronic band structure of the S1 (left) and S5 (right) graphene systems along the high symmetry \textbf{k}-points of the hexagonal BZ.} 
\end{center}
\end{figure*}

We show in Figure \ref{fig:elec_band_551} the electronic band structure of the S1 and S5 systems.

Consistent with the features observed in the eDoS (Figure \ref{fig:elec_dos_551}), changes in the dispersion characteristics between the two systems are evident. In both cases, the valence band maximum and the conduction band minimum are located at the high-symmetry K-point (the Dirac point); however, the linear feature is primarily observed along the high-symmetry \textbf{M}-\textbf{K}  segment. Both systems also exhibit flatter bands along other high-symmetry segments. However, in the case of S5, the combination of flat and dispersive electronic band structures lies closer to the Fermi energy \cite{PhysRevMaterials.9.L032401,doi:10.1021/acs.jpcc.4c08588,MADSEN2018140}, with the bands becoming flatter along the $\Gamma$–\textbf{M} direction. Along the \textbf{K}–$\Gamma$ segment, the dispersion becomes more parabolic as it approaches $E_F$ for S5, indicating a less pronounced linear character compared to the S1 system. This effect differs slightly from what has been reported for other structures \cite{PhysRevMaterials.9.L032401,doi:10.1021/acs.jpcc.4c08588,MADSEN2018140}, where the band remains flat along the direction closer to the Fermi energy and dispersive along other directions.

We have also computed the S5 electronic dispersion, when applying other external perturbation, such as an electric field along the $z$-direction. We observe that a small direct gap emerges with a value of 0.076 eV. We may even observe that the band edges become slightly parabolic, as opposed to the linear dispersion when no electric bias is applied, therefore slightly altering the qualitative effective masses of the charge carriers (Figure \ref{fig:elec_band_551_Bias} and more details are presented in Section \ref{electric_bias}).

\subsection{Charge Density Distribution}

\begin{figure*}[!]
\begin{center}
\includegraphics[width=14cm]{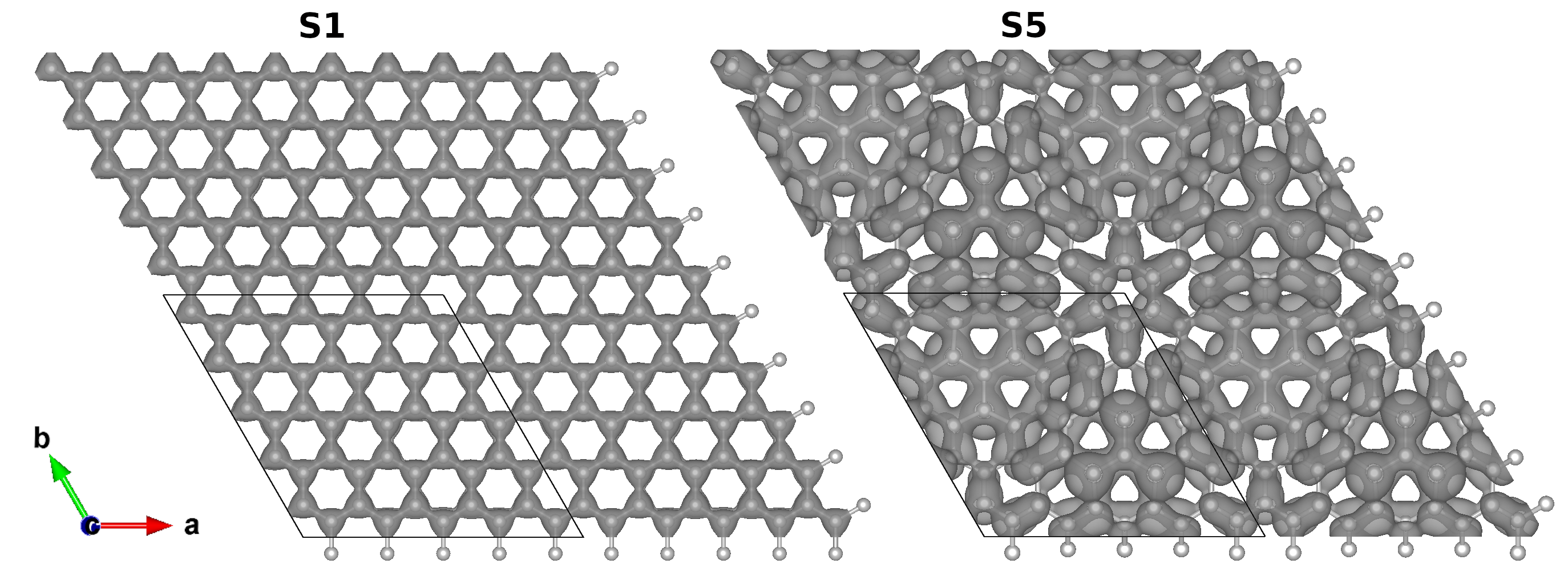}
\caption{\label{fig:parch}
Isosurfaces of the S1 (left) and S5 (right) enlarged supercells, for better visualization of the charge distribution patterns. Isosurface levels were set to 5$^{-6}$ e/\AA$^3$.} 
\end{center}
\end{figure*}

In Figure \ref{fig:parch} we show the isosurfaces computed around the Fermi energy for the S1 and S5 graphene cells.

For the S1 system, where the deformation is very mild, we observe a sp$^2$ type charge distribution, and coherent with a pristine hexagonal 2D graphene system. In contrast, differences are evident for the cell with greater deformation, where the charge distribution is no longer uniform. Specifically, on the concave regions of the curved surface, the charge distribution resembles that of an sp$^3$ hybridized configuration. This observation implies that curvature introduces localized charge distortions and disrupts the delocalized $\pi$-electron system. These sp$^3$-like regions may act as charge traps or scattering centers, thereby potentially modifying the value of $\kappa_L$ (details in Subsection \ref{kL}).

\subsection{Dynamical Stability}
%%%%%%%%%%%%%%%%%%%%%%%%%%%%%%%%%%%%

Energetic stability is a necessary, but not a sufficient condition for a structure to be synthesizable. One should also analyze the dynamical stability of the system, by analyzing the phonon (vibrational) properties. If imaginary frequencies are evident (usually represented by negative frequencies), this would indicate that the system is at a potential-energy maximum (transient state), undergoing a phase transition, and thus cannot be kinetically stable at the given perturbed condition \cite{ProcCambridgePhilosSoc36.160.1940,Dove.IntLattDyn,Dove.StrutDyn,AmMin.82.213.1997, BullMaterSci.1.129.1979, PhysRevLett.111.025503.2013}.

In suspended graphene the carbon atoms can oscillate along the out-of-plane direction leading to low-frequency flexural modes, with a quadratic dispersion relation \cite{OCHOA2012963, PhysRevB.88.115418}. However changes occur to this mode, when perturbations are induced.

\begin{figure*}
\begin{center}
\includegraphics[width=14cm]{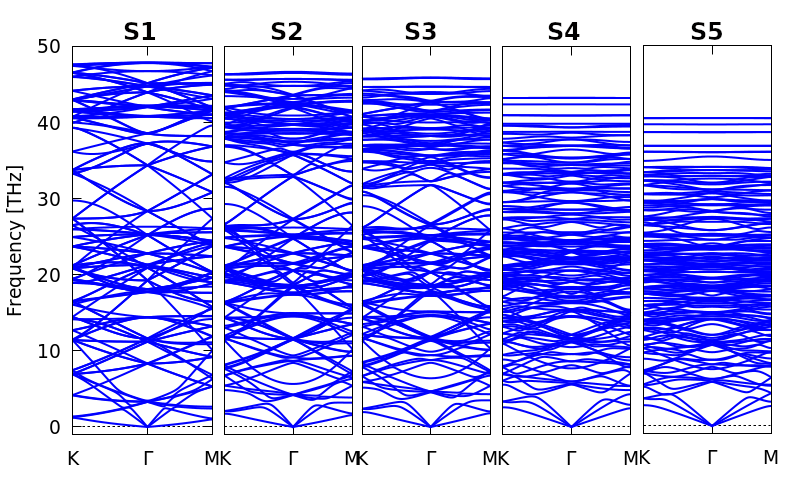}
\caption{\label{fig:band_551}
Phonon band dispersion of the five 5$\times$5$\times$1 deformed cells with different strain values.} 
\end{center}
\end{figure*}

In this section, we examine the phonon properties of the various cells as a function of curvature amplitude. Figure \ref{fig:band_551} illustrates the evolution of the flexural-acoustic (ZA) mode — the lower-frequency acoustic mode — which tends toward a linear dispersion behavior as the strain (curvature) increases. The behavior of the ZA mode exhibits 3D-like characteristics, analogous to that observed when graphene is adsorbed onto a substrate, where the mode hybridizes with the substrate itself \cite{PhysRevB.88.115418}. 

The parabolic character of the ZA mode in pristine graphene arises from the combined effects of the bending rigidity and elastic stiffness. The quadratic dispersion law serves as a signature of a true 2D crystal, wherein no restoring force exists for the out-of-plane bending at very long wavelengths, rendering the phonon soft. When the ZA mode becomes linear at low $q$, the long-wavelength flexural phonons propagate analogously to in-plane acoustic modes. In this regime, and if graphene is permanently curved (as in the S5 system), the effective strain induced by curvature introduces a restoring force that opposes the out-of-plane bending, thereby lifting the quadratic law. Consequently, the system behaves like a corrugated sheet in which the flexural modes are hardened.

Furthermore, as the degree of deformation increases, we observe a widening of the gap between the three acoustic modes and the optical modes. Additionally, the frequencies of the acoustic branches at the \textbf{K} and \textbf{M} \textbf{q}-points increase, indicating a hardening of these modes.

At the high-symmetry \textbf{K}-point of the S1 cell, a degeneracy is observed between the flexural acoustic mode and the lowest-frequency optical mode, both of which exhibit relatively low magnitudes. In contrast, for the S5 system, the acoustic modes all converge to a similar frequency range of approximately 5 THz, becoming clearly separated from the optical modes.
 
Moreover, it has been studied through Raman spectroscopy \cite{PhysRevLett.123.135501}, that unsupported monolayer graphene, when considered under pressure, shows an out-of-plane stiffness consistent with that of graphite. The authors recognized that the electron orbitals react to strain, resisting compression along the $x$-crystallographic axis; whereby the elastic constant, c33, would be defined by the resistance of the $p$ orbitals to compression. Such behavior is also consistent with what we observe form the isosurfaces plot of our deformed graphene sheet (Figure \ref{fig:parch}).

When inducing strain above a$_0$-0.1175 \%, imaginary phonons start to emergence, which are related to dynamical instabilities when such a magnitude of strain is applied. However, this feature can also be due to anharmonic effects taking over - and therefore the employed phonon harmonic approximation may break down. 

\subsection{Lattice Thermal Conductivity}\label{kL}

Several factors influence heat transport in graphene, including defects, grain boundaries, isotopic composition, strain, substrate interactions. These introduce phonon scattering sites that can either suppress or enhance heat conduction depending on the specific engineering approach \cite{c11030046}.

\begin{figure}
\begin{center}
\includegraphics[width=8cm]{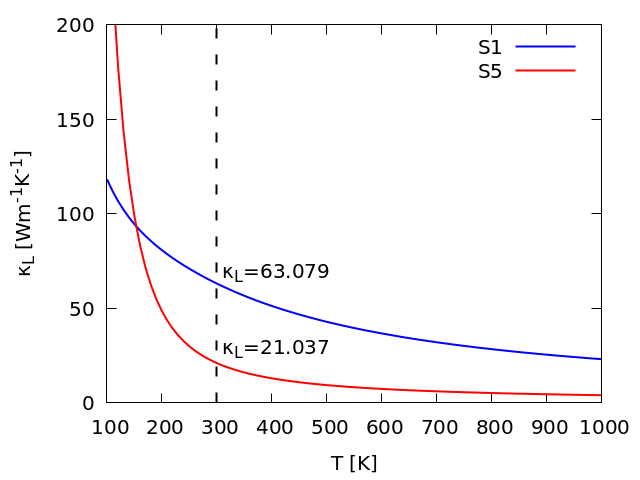}
\caption{\label{fig:thermal}
In-plane lattice thermal conductivity of the 5$\times$5$\times$1 cell of deformed graphene sheet of two different structures, the S1 and S5 (lower and higher deformation, respectively).} 
\end{center}
\end{figure}

We therefore compute the in-plane lattice thermal conductivity for two of the studied systems - the S1, with lowest stain-induced deformation; and for S5, with largest strain-induced deformation (Figure \ref{fig:thermal}).

We observe that the system with the highest deformation amplitude exhibits a lower magnitude of $\kappa_L$ compared to the system with the lowest deformation. At room temperature, the values of $\kappa_L$ are 21.04 Wm$^{-1}$K$^{-1}$ and 63.08 Wm$^{-1}$K$^{-1}$, respectively. It should be noted that for monolayer graphene, $\kappa_L$ is typically found to range between 1300 and 2200 Wm$^{-1}$K$^{-1}$ \cite{PhysRevB.108.L121412, PhysRevB.80.033406}, values not substantially different from those of bilayer graphene. Our results suggest that applying even a small amount of strain to the two-dimensional lattice can considerably reduce the value of $\kappa_L$. Moreover, as temperature increases, $\kappa_L$ for the S5 system remains lower than that for S1. Comparing this behavior with the phonon dispersion (Figure \ref{fig:band_551}), the out-of-plane flexural acoustic phonons provide the dominant contribution to $\kappa_L$, particularly as these transition from a parabolic to a linear dispersion characteristic. In freestanding graphene, the quadratic ZA mode contributes strongly to thermal resistance through Umklapp scattering \cite{PhysRevB.89.235422}. Conversely, linear ZA modes scatter less efficiently, which would typically increase thermal conductivity. However, our findings indicate that curvature-induced strain modifies this picture, leading to a net reduction in $\kappa_L$ despite the linearization of the ZA mode.

%For other well-known 3D thermoelectrics, the low-frequency optical phonon modes increase the available phase space of the phonon-phonon scattering of heat-carrying acoustic phonons, consequently lowering the lattice thermal conductivity values. Such feature occurs for As$_2$Te$_3$ \cite{PhysRevB.104.024103}, Cd$_3$As$_2$\cite{PhysRevResearch.1.033101},  PbTe \cite{PhysRevLett.112.175501} and SnSe \cite{NatPhys11_1063}. However, for the 2D layer system, we find that this interpretation is not valid, since we observe that as soon as the optical modes harden in frequency, $\kappa_L$ eventually drops. 

\begin{figure*}
\begin{center}
\includegraphics[width=14cm]{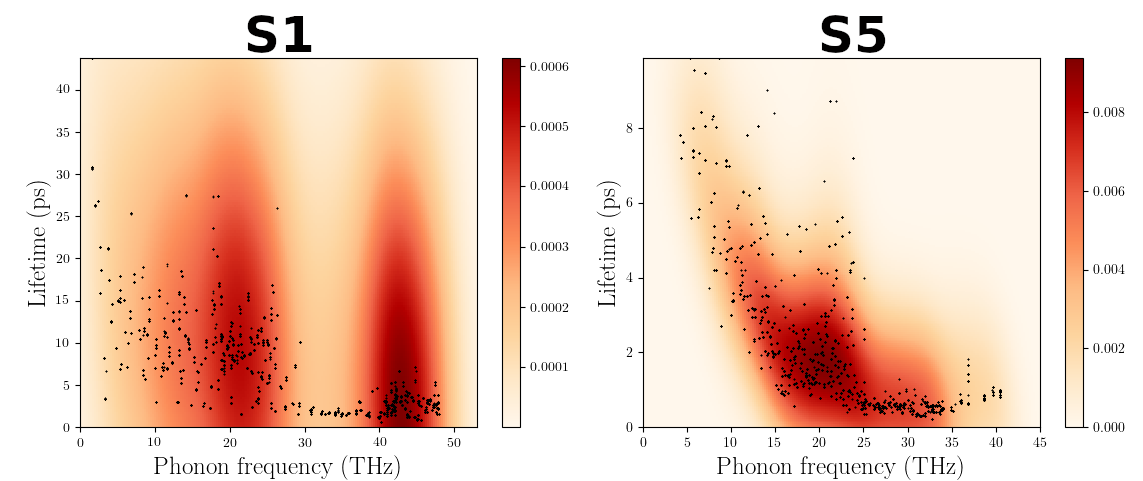}
\caption{\label{fig:lifetime}
Calculated phonon lifetimes of the S1 (left) and S5 (right) systems at 300 K. The color shades represent the phonon density, consequently darker shades refer to higher phonon densities.} 
\end{center}
\end{figure*}

%phonon lifetimes
The phonon lifetime determines the magnitude of $\kappa_L$. We have computed the phonon lifetimes of the S1  and S5 cells - Figure \ref{fig:lifetime}.

For the S1 surface, the phonon lifetime density is predominantly concentrated in the 20–25 THz and 40–47 THz regions, with the latter range exhibiting greater significance and featuring high lifetimes of approximately 30 ps. In contrast, for the S5 surface, the higher phonon density is concentrated within a single broad packet centered approximately between 15 and 25 THz. The S5 system exhibits very low phonon lifetimes, which accounts for its lower lattice thermal conductivity compared to S1 (21.04 Wm$^{-1}$K$^{-1}$ \textit{vs} 63.08 Wm$^{-1}$K$^{-1}$, respectively). 

%In a comprehensive theoretical study of thermal transport in diamond, graphene, and graphite \cite{PhysRevB.89.155426}, it was shown that the sharp peaks in the phDoS (due to the VHS) are directly related the phonon lifetime spectra. This anti-correlation between phDoS and phonon lifetime is a hallmark of the VHS scattering mechanism.

%Another study performed on boron arsenide \cite{Tadano_2014}, the authors found an unusually high thermal conductivity partly because the phonon dispersions were designed to avoid VHS overlaps. The acoustic and optical branches were well-separated, and the acoustic branches were highly non-dispersive (linear), minimizing flattening of the transverse acoustic (TA) branch along the $\Gamma$-K direction creates a VHS. The authors explicitly stated that such a feature was expected to cause strong anharmonic scattering and low thermal conductivity, linking the dispersion feature directly to the intrinsic properties of the material.

\section{Conclusions}

We have demonstrated the possibility of tuning the electronic, vibrational, and lattice thermal conductivity properties of monolayer graphene by inducing x-y strain deformations that create topological curvature on the sheet surface. The induced curvature with the defined strain constraint energetically stabilizes the systems, enabling these to become more stable than flat graphene under the same applied strain magnitude. The curved systems are dynamically stable within the studied range of strains/curvatures, as confirmed by the phonon dispersion calculations that show no imaginary frequencies (until exceeding a$_0$-0.1175\%). 

As deformation increases, the VHS in the eDoS approach the Fermi energy. The highly curved S5 system exhibits coexisting flat and linear dispersions close to the Fermi level — a promising feature for thermoelectric applications.

The ZA mode transitions from a quadratic to a linear dispersion under increasing strain/curvature. This mimics 3D behavior, stiffens the out-of-plane phonon modes, and suppresses the characteristic 2D quadratic dispersion — promoting 3D-like phonon dynamics with direct consequences for thermal transport and mechanical stability.

The increase of phonon scattering (due to curvature-induced strain and sp$^3$-like charge localization) decreases the value of $\kappa_L$. At room temperature, $\kappa_L$ drops from 63.08 Wm$^{-1}$K$^{-1}$ (S1, low deformation) to 21.04 Wm$^{-1}$K$^{-1}$ (S5, high deformation). This demonstrates that it is possible to tune $\kappa_L$ by applying x-y strain and inducing different topological curvatures.

The integration of flat band physics into thermoelectric materials research represents a paradigm shift — moving beyond simple heavy-band semiconductors toward correlated, topological, and engineered strain-induced systems. The key is not only to obtain a high electronic density of states, but to achieve it within an electronic band structure that still permits good carrier transport.

Future directions will focus on integrating strain-engineered graphene insights into functional devices, exploring different 2D strain architectures, and leveraging machine learning for predictive strain engineering. Curved graphene surfaces introduce unique 3D features that enhance surface area for catalysis, enable tuning of conductivity through electron localization, and allow novel designs for energy storage, sensors, and biomedical applications.

%%%%%%%%%%%%%%%%%%%%%%%%%%%%%%%%%%%%%%%%%%%%%%%%%%%%%%%%%%%%%%%%%%%%%
%% The "Acknowledgement" section can be given in all manuscript
%% classes.  This should be given within the "acknowledgement"
%% environment, which will make the correct section or running title.
%%%%%%%%%%%%%%%%%%%%%%%%%%%%%%%%%%%%%%%%%%%%%%%%%%%%%%%%%%%%%%%%%%%%%
\section{Acknowledgement}

This research was supported by the Por\-tu\-guese Foundation for Science and Technology (FCT) with 2022.00082.CEECIND (doi.org/\-10.54499/\-2022.00082.CEECIND/\-CP1719/\-CT0001) grant; UIDP/04968/2020 (doi.org/\-10.54499/\-UIDP/\-04968/\-2020) and UIDB/04968/2020 (doi.org/\-10.54499/\-UIDB/\-04968/\-2020); and the FCT - Mobility outgoing with Ref. FCT/\-Mobility/\-1303210959/\-2024-25 projects. ELdS further acknowledges the High Performance Computing Chair - a R\&D infrastructure (based at the University of Évora; PI: M. Avillez), endorsed by Hewlett Packard Enterprise, and involving a consortium of higher education institutions, research centers, enterprises, and public/private organizations. Additionally, the authors acknowledge the computer resources from the FCT project 2022.15832.CPCA.A2 to access Oblivion (hosted by High Performance Computing Univ. Évora), 2025.00082.CPCA.A3 to access MareNostrum5 (hosted by BSC) and the the use of ARCHER2 UK National Supercomputing Service through membership of the UK's HEC Materials Chemistry Consortium (HEC MCC), which is funded by the EPSRC (EP/X035859/1). This publication is based upon work from the COST Action FuSe, CA24101, supported by the European Cooperation in Science and Technology).
\clearpage

%%%%%%%%%%%%%%%%%%%%%%%%%%%%%%%%%%%%%%%%%%%%%%%%%%%%%%%%%%%%%%%%%%%%%
%% The same is true for Supporting Information, which should use the
%% suppinfo environment.
%%%%%%%%%%%%%%%%%%%%%%%%%%%%%%%%%%%%%%%%%%%%%%%%%%%%%%%%%%%%%%%%%%%%%
\section{Appendix}

\subsection{Structural Stability}

\begin{figure*}[!]
\begin{center}
\includegraphics[width=14cm]{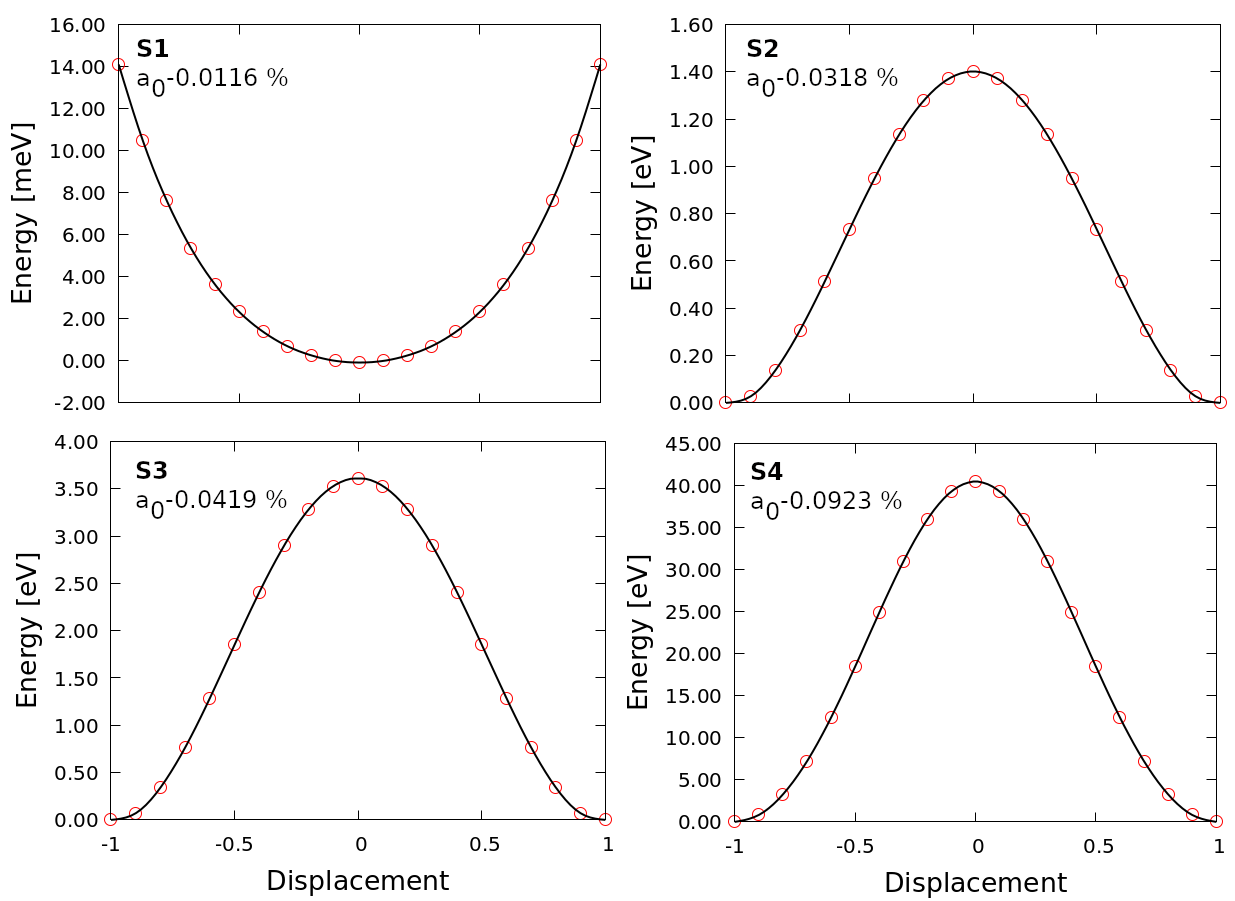}
\caption{\label{fig:potential_well_4}
Potential energy surfaces of the four graphene systems for different induced-strain. } 
\end{center}
\end{figure*}

We present the potential energy surfaces for the four graphene systems with varying deformation amplitudes, ranging from S1 to S4 (Figure \ref{fig:potential_well_4}), following a procedure analogous to that described in Subsection \ref{energy_stability}.

With the exception of the S1 system, for which the flat 2D graphene layer represents the most stable configuration, the remaining systems exhibit an energetic preference for the deformed state when lattice constraints are applied. As lattice strain increases, the potential energy surface becomes progressively higher, rendering the spontaneous transition to the flat monolayer configuration increasingly unfavorable.
 
\subsection{C-C Bonding Length}

\begin{figure*}[!]
\begin{center}
\includegraphics[width=12cm]{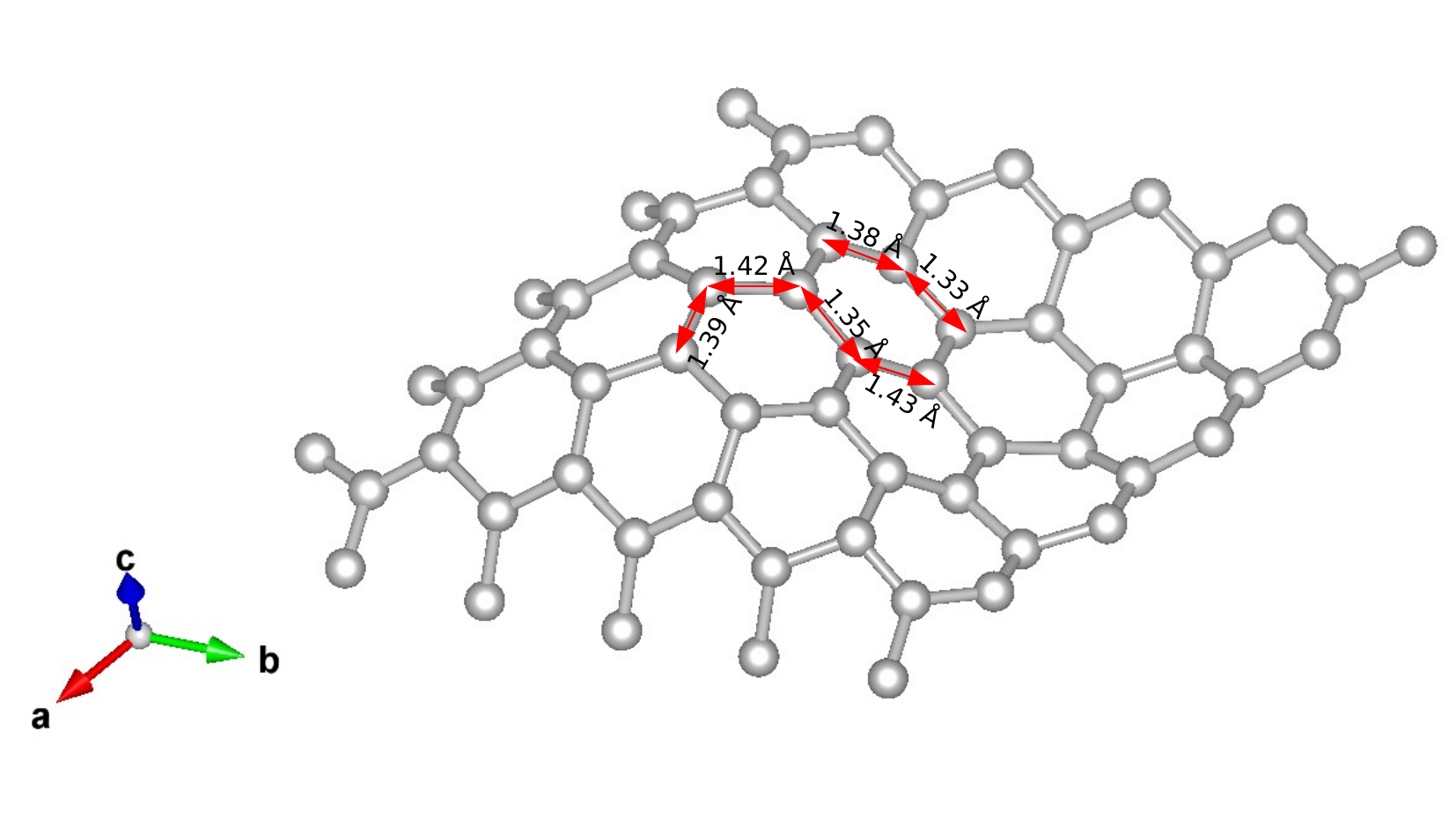}
\caption{\label{fig:CC_bond}
C-C bond length differences of the S5 graphene system.} 
\end{center}
\end{figure*}

It is well established that unstrained, flat graphene exhibits a C–C bond length of 1.42~\AA. In our PBEsol calculations, the S1 system, which possesses the lowest deformation, displays a C–C bond length of 1.40~\AA. In contrast, for the S5 system (characterized by the highest curvature), the C–C bond lengths vary across the hexagonal sublattice as a function of the deformation position, ranging between 1.33~\AA and 1.43~\AA, as shown in Figure \ref{fig:CC_bond}.

\subsection{Applied Electric Bias}
\label{electric_bias}

\begin{figure*}[!]
\begin{center}
\includegraphics[width=12cm]{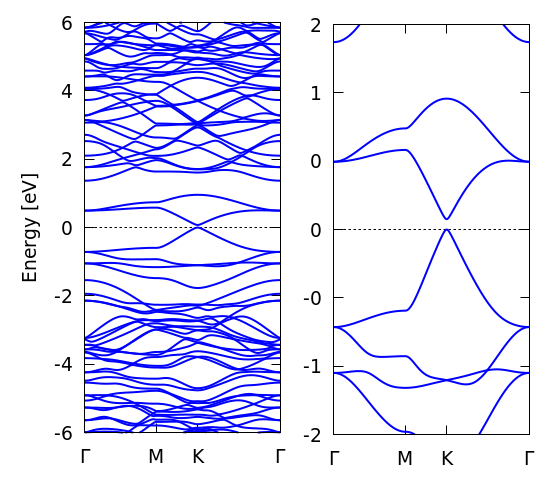}
\caption{\label{fig:elec_band_551_Bias}
Electronic band structure of the S5 graphene system along the high symmetry \textbf{K}-points of the hexagonal BZ, without (left) and with (right) applied electric bias of 0.1 eV/\AA.} 
\end{center}
\end{figure*}

Upon applying a finite electric field along the crystallographic $z$-direction, the small perturbation destabilizes the degenerate spectrum, thereby generating a small single-electron gap - Figure \ref{fig:elec_band_551_Bias}. 

Such a behavior is similar to that observed in bilayer graphene \cite{SILVA2020373}, where the two layers are subjected to inequivalent potentials under an applied electric bias. This effect breaks inversion symmetry, leading to the opening of a single-electron gap at the \textbf{K}-point, which can be tuned to energies as high as the mid-infrared range (approximately 300 meV). However, for our monolayer structure, we observe that the gap opens up until 76 meV when applying an electric bias of 0.1 eV/\AA. Furthermore, since spontaneous breaking of translation symmetry occurs, it results in charge separation between the inequivalent hexagonal 'sublattices' accompanied by spatial in-plane charge inhomogeneities \cite{SILVA2020373}.

\clearpage
%\bibliography{biblo}
%apsrev4-2.bst 2019-01-14 (MD) hand-edited version of apsrev4-1.bst
%Control: key (0)
%Control: author (8) initials jnrlst
%Control: editor formatted (1) identically to author
%Control: production of article title (0) allowed
%Control: page (0) single
%Control: year (1) truncated
%Control: production of eprint (0) enabled
%

\end{document}